\documentclass[a4paper,11pt]{article}
\usepackage{pos}

\title{Baryogenesis through leptogenesis in the minimal flipped $SU(5)$  with radiative seesaw}

\author*[a]{Michal Malinsk\'{y}}
\author[b]{Renato Fonseca}
\author[a]{V\'{a}clav Mi\v{r}\'{a}tsk\'{y}}
\author[a]{Martin Zdr\'{a}hal}

\affiliation[a]{Institute of Particle and Nuclear Physics,\\
  Faculty of Mathematics and Physics,
  Charles University in Prague, V Hole\v{s}ovi\v{c}k\'ach 2,
  180 00 Praha 8, Czech Republic}
\affiliation[b]{Departamento de Física Teórica y del Cosmos,\\ Universidad de Granada, Campus de Fuentenueva, E-18071 Granada, Spain}
\emailAdd{michal.malinsky@matfyz.cuni.cz}

\abstract{The minimal flipped $SU(5)$ unification with the right-handed neutrino Majorana mass scale generated as a two-loop effect is arguably one of the most constrained models of perturbative baryon and lepton number violation currently on the market. This is namely due to its very simple scalar sector structure which, subject to perturbativity and non-tachyonicity bounds, leads to the emergence of characteristic flavour patterns providing interesting insights into proton decay phenomenology, neutrino physics etc.    

In this contribution, we discuss the potential impact of the baryon asymmetry of the Universe as an additional requirement imposed onto the already strongly constrained flavour structure of the model, focusing on thermal leptogenesis as its hypothetical primary source in this framework. Remarkably, this single extra condition leads, among other things, to a very interesting upper limit on the absolute neutrino mass scale which, in turn, renders the model potentially testable in the existing beta-decay experiments such as KATRIN.     
}

\FullConference{Proceedings of the Corfu Summer Institute 2024 "School and Workshops on Elementary Particle Physics and Gravity" (CORFU2024)\
12 - 26 May, and 25 August - 27 September, 2024, 
Corfu, Greece\\}

\begin{document}
\maketitle

\section{The flipped $SU(5)$ framework}
Even after four decades since its conception, the ``flipped'' $SU(5)$  model characterised by a non-trivial identification of the SM hypercharge within the Cartan algebra of the underlying $SU(5)\times U(1)_X$ gauge structure~\cite{Barr:1981qv,Derendinger:1983aj} (cf. Table~\ref{table1}) is still a rather popular (partially) unified extension of the Standard Model (SM). Besides admitting a very simple spontaneous symmetry breaking pattern\footnote{To this end it is perhaps worth noting that unlike in the Georgi-Glashow context, a single-step breaking in the flipped $SU(5)$ is right away consistent with the SM gauge coupling unification constraints as, in the current context, it effectively concerns only the two non-abelian ones.} governed by a smaller set of scalar fields\footnote{In principle, the $SU(5)\times U(1)$ gauge structure can be broken down to the $SU(3)\times SU(2)\times U(1)$ of the SM and later to $SU(3)\times U(1)$ of QCD $\times$ QED by as little as one copy of an SU(5) scalar ten-plet $({\bf 10},+1)$ and one copy of a five-plet $({\bf 5},-2)$; however, the minimal potentially realistic variant contemplated in this review requires two copies of the latter, cf.~\cite{ArbelaezRodriguez:2013kxw}.} than the ``standard'' $SU(5)$ GUT \`{a} la Georgi and Glashow~\cite{Georgi:1974sy}, it makes the right-handed neutrinos (RHNs) a very natural (in fact, inevitable) part of the play. Moreover, it correlates their Dirac mass matrix to that of the up-type quarks 
\begin{equation}
\label{MnuDirac}
M_\nu^D=M_u^T\,,
\end{equation}
thus avoiding the somewhat unpleasant degeneracy between the masses of down-type quarks and charged leptons encountered in the simplest variants of the G-G model. 
\begin{center}
\begin{table}[h]
\begin{tabular}{|cc|c|c|c|}
\hline
$SU(5)\times U(1)_X$ & SM & $T_{24}$ & $Y=\tfrac{1}{5}(X-T_{24})$ & $B-L=\tfrac{1}{5}(X+4\,T_{24})$\\
\hline
$({\bf 1},+5)_M$ & $e^c$ & 0 &  $+1$ & $+1$\\
$(\overline{\bf 5},-3)_M$ & $u^c, L$ & $+\tfrac{1}{3}$, $-\tfrac{1}{2}$ & $-\tfrac{2}{3}$, $-\tfrac{1}{2}$ & $-\tfrac{1}{3}$, $-1$\\
$({\bf 10},+1)_M$ & $d^c, Q,\nu^c$ & $-\tfrac{2}{3}$, $+\tfrac{1}{6}$, $+1$ & $+\tfrac{1}{3}$, $+\tfrac{1}{6}$, $0$ & $-\tfrac{1}{3}$, $+\tfrac{1}{3}$, $+1$\\
\hline
\end{tabular}
\caption{\label{table1}Quantum numbers of the (three generations of) matter fields in the flipped $SU(5)$ context. The ``flipping'' in the matter spectrum (as compared to the ``standard'' Georgi-Glashow $SU(5)$  field assignment) concerns the SM right-handed matter fields, namely $d^c\leftrightarrow u^c$ and $e^c\leftrightarrow \nu^c$. Note that in the symmetric phase $B-L$ is a part of gauge symmetry independent of (though not orthogonal to) the SM hypercharge $Y$ and, due to the absence of a SM  singlet in the relevant ten-plet (see the lower-right entry above) it is spontaneously broken once the VEV of the $({\bf 10},+1)$ scalar is engaged.} 
\end{table}
\end{center}
In this respect, the flipped $SU(5)$ can be seen as a highly economical alternative to even the most simple variants of the ``classical'' Grand unified theoreis (GUTs) such as the minimal $SO(10)$ model (cf.~\cite{Bertolini:2009es,Jarkovska:2021jvw}), with a potential to address the same type of questions concerning baryon and lepton number violation effects (like those related to proton longevity or Majorana neutrino phenomenology); for details see e.g.~\cite{Dorsner:2004jj,Dorsner:2004xx} and references therein.    
\subsection{The minimal flipped $SU(5)$ model with radiative RHN mass generation\label{sect:minimalflippedSU5}}
The natural presence of RHNs and the strong correlation between the associated $M_\nu^D$ and $M_u$  (\ref{MnuDirac}) characteristic to essentially all flipped $SU(5)$ settings, clearly calls for a type-I seesaw mechanism implementation.  
Remarkably, with RHNs residing in $({\bf 10},+1)$'s of $SU(5)\times U(1)_X†$, a tree-level generation of their Majorana masses requires an appropriate irreducible representation from the symmetric product of two ${\bf 10}$'s of $SU(5)$, namely, a $({\bf 50},-2)$ scalar multiplet. This option, however, makes the model suffer from similar drawbacks as its ``standard'' $SU(5)$ counterpart, namely, a sequestered RHN sector with little to no correlation of thus generated Majorana mass term ($M_\nu^M$) structure to other Yukawa couplings in the game. 
\begin{figure}[tb]
\centerline{%
\includegraphics[width=4cm]{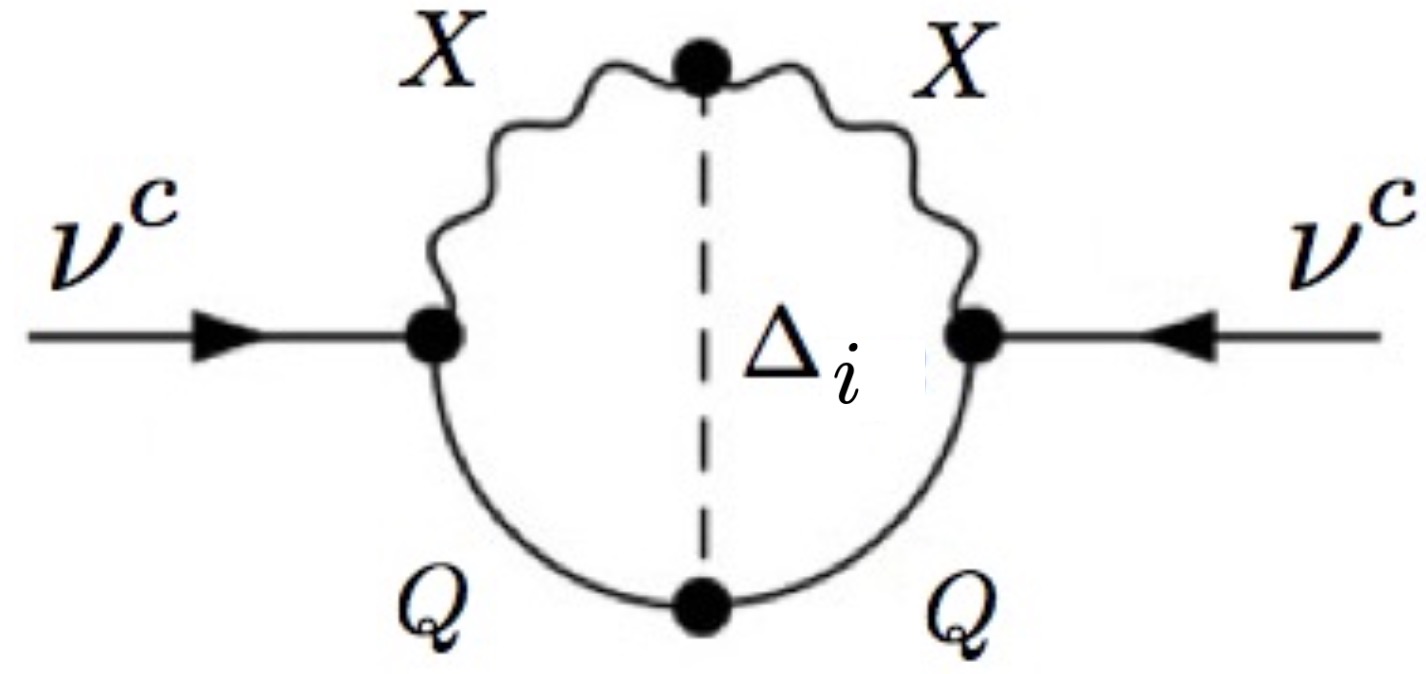}\hspace{1cm}
\includegraphics[width=4cm]{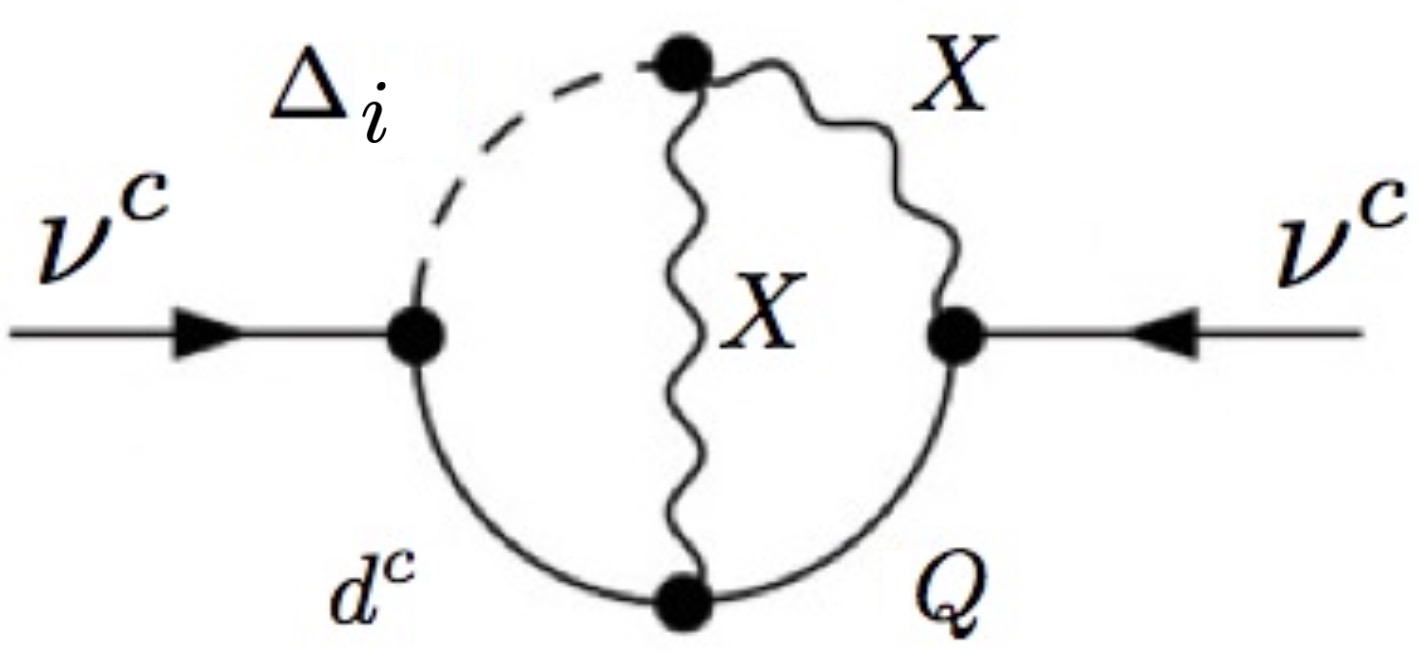}}
\caption{Radiative generation of RHN Majorana mass in the minimal flipped $SU(5)$ model~\cite{ArbelaezRodriguez:2013kxw}. $X$ and $\Delta_i$ denote the high-scale vector bosons and scalar $SU(3)$ triplets, respectively.}
\label{figuregraphs}
\end{figure}

To this end, a far more attractive alternative addressing all these concerns is a  two-loop mechanism identified in~\cite{ArbelaezRodriguez:2013kxw,Harries:2018tld} (see also~\cite{Leontaris:1991mq}), in fact a simplified variant of the Witten's proposal~\cite{Witten:1979nr} originally made in the $SO(10)$ context; see Fig.~\ref{figuregraphs}. Notably, it requires no additional scalars beyond those already present and, as a bonus, provides a rationale for the anticipated gap between the unification and seesaw scales. Moreover, the structure of $M_\nu^M$ in this case is correlated to the Yukawa couplings driving the charged-fermion sector.  This, together with the technical requirements of perturbativity and non-tachyonicity of the scalar spectrum, leads to a number of interesting patterns in $B$- and/or $L$-violating observables such as limits on the absolute (Majorana) neutrino mass scale or features in two-body proton decay branching ratios, see~\cite{ArbelaezRodriguez:2013kxw}. 

As for the latter, the central structure ruling all the two-body $p$-decay channels is a unitary matrix $U_\nu$ diagonalising the light neutrino mass matrix in the basis in which $M_\nu^D$ is diagonal, for instance
\begin{align}
\label{form1}
\Gamma(p\to \pi^0 \mu^+)& =\frac{1}{2}|(V_{CKM})_{11}|^2|(V_{PMNS}U_\nu)_{21}|^2 \Gamma(p\to \pi^+\overline{\nu})\,.
\end{align}
For a given neutrino hierarchy, the shape of the $U_\nu$ matrix, together with the mass of the lightest active neutrino, defines the region of the  parameter space of our main interest where all the basic low-energy flavour constrains (namely, those related to the known quark and lepton masses and mixing parameters) are implicitly satisfied.  
\section{Baryogenesis through leptogenesis in the minimal flipped $SU(5)$ model}
The question addressed in the recent study~\cite{Fonseca:2023per} is basically how large a part of this physical parameter space domain is further cut out -- and what extra features potentially emerge -- if one demands a full account of the baryon asymmetry of the Universe in the current model. With a strongly constrained heavy neutrino spectrum at hand, one naturally expects a particularly good grip on thermal leptogenesis~\cite{Fukugita:1986hr} which can be expected to be among primary sources of dynamically generated net $B-L$ in this scenario.      

\subsection{Dynamical $B-L$ generation in flipped $SU(5)$ models}
As formal descendants of the $SO(10)$ GUTs the flipped $SU(5)$ models provide a clear identification of $B-L$ as a linear combination of the $SU(5)$ and $U(1)_X$  generators, see Table~\ref{table1}.
In settings like~\cite{King:2024gsd} where the high-energy symmetry is first broken by $({\bf 24},0)$, $B-L$ is retained as an unbroken gauge symmetry generator down to a lower scale where the rank is eventually reduced, with potentially interesting signals emerging (like, e.g., cosmic-string produced stochastic gravitational wave spectrum component etc.) at some intermediate scale. At the same time, since no $B-L$ asymmetry can be generated at or close to the unification scale, thermal leptogenesis is the most natural source of the net $B-L$ generated in later stages. 

In general, this does not need to be the case in the ``genuine'' flipped $SU(5)$ model under consideration in which the VEV of the $({\bf 10},+1)$ scalar breaks $U(1)_X$ at the unification scale along with the $SU(5)$ symmetry. In such situations a net $B-L$ may be first produced well above the seesaw scale and one should not a-priori ignore it; rather than that, it should be taken as an initial condition for the subsequent thermal leptogenesis stage. However, the assumption of the dominance of the latter  can still be justified if, for instance, a strong washout precedes the out-of-equilibrium decays of RHNs, or if there is a reason for a suppression of the CP asymmetry in the decays of the heavy colour triplets  (for example, their strongly hierarchical Yukawas). In what follows we shall simply assume that this is indeed the case and, hence, thermal leptogenesis happens to be the primary source of the net $B-L$ also in the current scenario.
      
\subsection{Leptogenesis in the minimal flipped $SU(5)$ model\label{leptogenesis}}
At first glance, one may expect that attaining a large-enough lepton asymmetry (and, thus, baryon-to-photon number density ratio $\eta_B$ in the ballpark of the experimental value of $6\times 10^{-10}$) may be rather non-trivial in the current setting, at least for the lightest active neutrino mass in the sensitivity domain of the ongoing $\beta$-decay experiments like KATRIN~\cite{KATRIN:2001ttj,KATRIN:2022ayy}. The main reason for this is the simple correlation between the light and heavy neturino mass spectra\footnote{Note that for a given hierarchy all three active nutrino masses are simple functions of a single parameter corresponding to the lightest of them (modulo uncertainties in determination of the two mass-squared differences measured in neutrino oscillation experiments).} in the form
\begin{equation}
\label{masses}
m_1m_2m_3M_1M_2M_3=m_u^2m_c^2m_t^2\sim 1.3\, {\rm GeV}^6\,,
\end{equation}
(with all quantities therein evaluated at the seesaw scale) which  stems from relation~(\ref{MnuDirac}). This, together with the two-loop suppression associated to the quantum nature of the seesaw scale in the current scenario, favours a relatively low-scale RHN spectrum, at a potential conflict with the basic Davidson-Ibarra (DI) limit~\cite{Davidson:2002qv} (at least as long as the heavy RHN masses $M_i$ happen to be hierarchical). From the opposite perspective, this also suggests several regimes in which the qualitative DI bound can be avoided -- the RHN spectrum does not need to be very hierarchical, the asymmetry may emerge from decays of the next-to-lightest rather than lightest RHN (assuming it is save from a subsequent strong washout) etc.   

\begin{figure}[tb]
\centerline{%
\includegraphics[width=8cm,height=7cm]{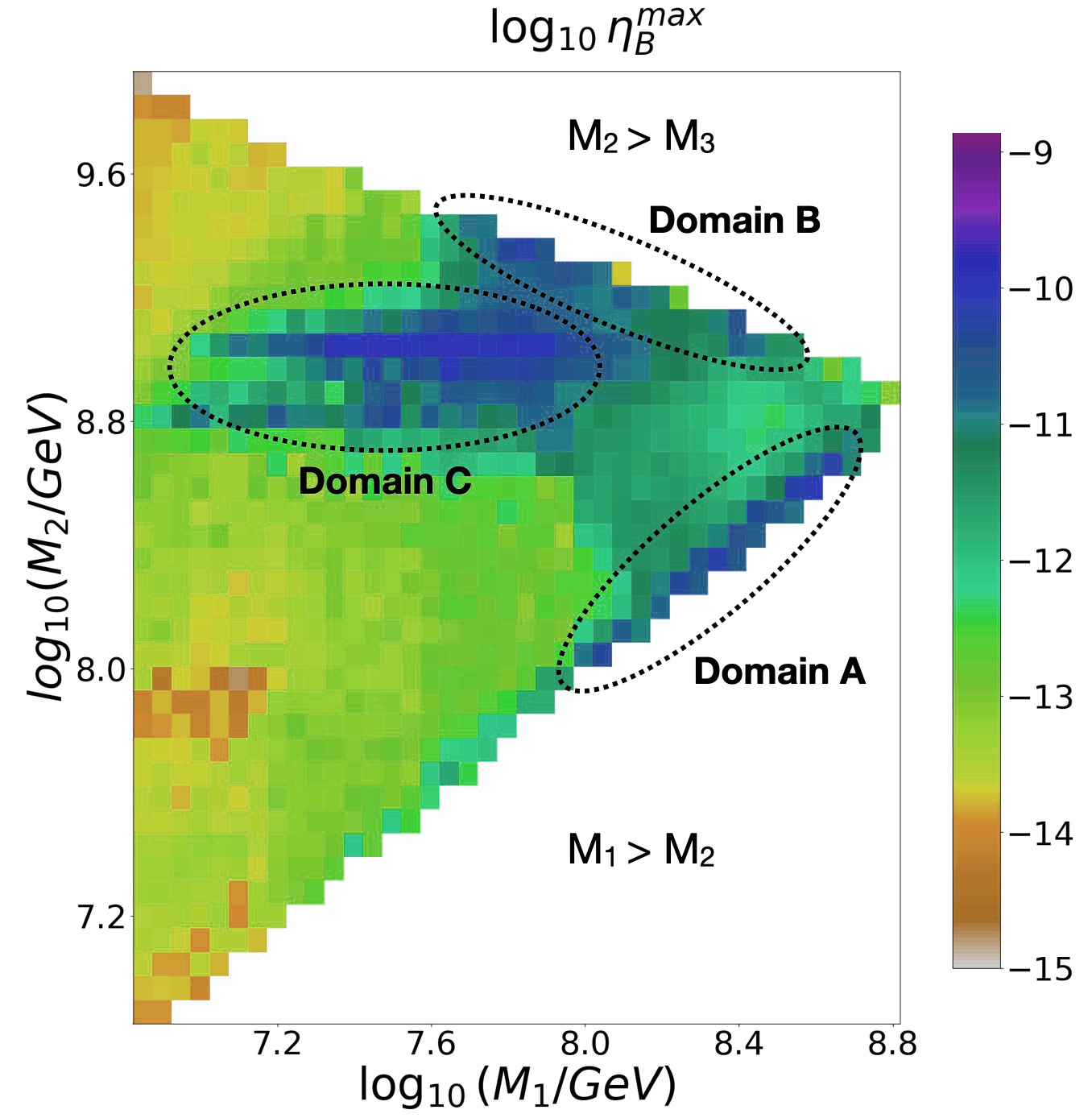}}
\caption{The maximum $\eta_B$ attainable in a sample scan of the minimal flipped $SU(5)$ model parameter space for a relatively compressed normally ordered active neutrino spectrum with $m_1$ around $3\times 10^{-2}$~eV. Due to the constraint~(\ref{masses}) the relevant domain has a triangular shape in the $\log -\log$ plane, with the rightmost apex corresponding to all $M_i$'s around $10^{8.8}$ GeV. It is worth noting that the plot represents a two-dimensional cut through a multidimensional parameter space which, besides the RHN masses, comprises several additional high-scale parameters such as extra phases (see Ref.~\cite{Fonseca:2023per}). Three qualitatively different domains where $\eta_B^{\rm max}$ peaks can be identified, see discussion in Sect.~\ref{leptogenesis}. Note also that the value $3\times 10^{-2}$~eV corresponds to the upper limit on $m_1$ for which the maximum attainable $\eta_B$ becomes first (albeit still marginally) compatible with the observed value $\eta_B^{\rm obs.}\sim 6\times 10^{-10}$; hence, $m_1 < 3\times 10^{-2}$~eV can be interpreted as a conservative upper limit on the mass of the lightest active neutrino in the current scenario.}
\label{figure1}
\end{figure}

Remarkably, this is exactly what has been observed in a dedicated analysis of Ref.~\cite{Fonseca:2023per} utilising the ULYSSES numerical package~\cite{Granelli:2020pim} to calculate $\eta_B$ for every point in an extensive parameter space scan of the current model, see Fig.~\ref{figure1}. Three qualitatively different domains in which $\eta_B$ can attain the desired level can be distinguished there:
\begin{enumerate}
\item
Domain A: In this parameter space area the asymmetry production is dominated by the decays of the lightest RH neutrino $N_1$ with a relatively suppressed washout; the DI limit is avoided by a relative proximity of $M_1$ and $M_2$ which, however, never reaches the limits of the resonant mode.
\item
Domain B: Here $N_2$ decays dominate the asymmetry production which is to a large degree protected from the subsequent $N_1$-driven washout by the suppression of  decoherence effects. In this case, the asymmetry production is partially enhanced due to the relative proximity of $M_2$ and $M_3$.    
\item
Domain C: In this region none of the asymmetries is enhanced due to a relatively hierarchical nature of the RHN spectrum (at least as compared to the previous two cases) and the asymmetry produced in the $N_2$ decays is kept alive only due to the strong suppression of the associated washout/decoherence effects. 
\end{enumerate} 
Remarkably enough, in all these cases, $\eta_B$ in the $6\times 10^{-10}$ ballpark can only be achieved for the mass of the lightest active neutrino below about $3\times 10^{-2}$~eV. In that respect, the situation depicted in Fig.~\ref{figure1} represents the limiting case on the verge of marginal compatibility with the observational data. This, translated to the upper limit on the $m_\beta$ parameter measured in the ongoing beta-decay experiments such as KATRIN, provides a testable prediction of the current scenario!   
 
\subsection{Implications for proton decay}
The upper limit on $m_1$, together with the constraints on the shape of the RHN spectrum and the additional high-scale mixing parameters (cf.~\cite{Fonseca:2023per}) as depicted in Fig.~\ref{figure1}, provides further constraints on the shape of the $U_\nu$ matrix driving proton decay amplitudes such as~(\ref{form1}), complementary to the aforementioned bounds from perturbativity and non-tachyonicity. The most intriguing of these features is the universal upper limit on the branching ratio ${\rm BR}(p^+\to \pi^0 \mu^+)$ which nowhere in the $\eta_B$-compatible part of the parameter space exceeds $30\%$. Interestingly, this bound gets even more strict for $m_1$ significantly below the $3\times 10^{-2}$~eV level (for instance, for $m_1\sim 10^{-3}$~eV one has ${\rm BR}(p^+\to \pi^0 \mu^+)\lesssim 10\%$ and so on).
     
\subsection{Conclusions}
Unified gauge models, especially in their simplest incarnations, represent very powerful frameworks for addressing some of the deepest mysteries of today's particle physics, especially when it comes to modelling the effects of baryon and/or lepton number non-conservation. In the current study, we have examined the very minimal $SU(5)\times U(1)$ setting with a ``flipped'' hypercharge assignment, in which the seesaw scale emerges as a quantum effect at two-loops, from the perspective of the baryon number problem of the Universe. It turns out that, contrary to the na\"\i ve expectation, the measured baryon-to-photon number density ratio can be accommodated in several regions of its parameter space, but only when the lightest active neutrino is not heavier than about $3\times 10^{-2}$~eV. Furthermore, a rather characteristic  pattern emerges in two-body proton decay branching ratios. This makes the minimal flipped $SU(5)$ a very physical theory which can be potentially testable in the current and next generation experimental facilities such as KATRIN, Hyper-Kamiokande or DUNE.                 

\section*{Acknowledgments}
This work has been supported by the FORTE project CZ.02.01.01/00/22\_008/0004632 co-funded by the EU and the Ministry of Education, Youth and Sports of the Czech Republic, and by the Charles University Research Center Grant No. UNCE/24/SCI/016. 


\end{document}